\documentclass[]{aa}
\usepackage{epsfig} 
\def\uka { \raisebox{-0.5ex} {\mbox{$\stackrel{<}{\scriptstyle 
\sim}$}}}
\def\uga { \raisebox{-0.5ex} {\mbox{$\stackrel{>}{\scriptstyle 
\sim}$}}}

\begin{document}
\thesaurus{3 (11.02.2, 13.07.2, 13.25.2)}
\title{X-ray/TeV-gamma-ray observations of several strong flares of Mkn 501 during 1997 and
implications}
\author{
H. Krawczynski\inst{1},
P.S. Coppi\inst{2},
T. Maccarone\inst{2},
F.A. Aharonian\inst{1},
}
\institute{Max Planck Institut f\"ur Kernphysik,
Postfach 103980, D-69029 Heidelberg, Germany \and
Yale University, P.O. Box 208101, New Haven, CT 06520-8101, USA}
\mail{Henric Krawczynski}
\offprints{Henric Krawczynski (email address: Henric.Krawczynski@mpi-hd.mpg.de)}
\date{Received 12 July 1999; accepted 27 October 1999}
\authorrunning{H. Krawczynski et al.}
\titlerunning {X-ray/TeV-gamma-ray observations of Mkn 501 during 1997 and implications}
\maketitle
\begin{abstract}
During more than 6 months in 1997, the BL Lac object Mkn 501 was in an exceptionally
bright state, both in the X-ray band and in the Very High Energy (VHE) band. 
In this paper we present a multiwavelength study of 
Mkn 501 during this extraordinary outburst.
We describe the analysis of a data base of X-ray observations acquired with the
pointed X-ray telescopes on board the {\it Rossi X-Ray Timing Explorer} (RXTE)
during April, May, and July, 1997.
We combine this data set with detailed VHE spectral information obtained from 
simultaneous or nearly simultaneous observations with the stereoscopic 
Cherenkov telescope system of HEGRA.
Several strong flares could  clearly be resolved in both energy bands, making it
possible to perform a detailed correlation analysis of the X-ray and 
VHE $\gamma$-ray flux levels and spectra.  
We interpret the results in the framework of a Synchrotron Self Compton 
models and study the constraints on the model parameters. 
We show that the emission mechanism produces an Inverse Compton spectrum 
with a significant curvature in the TeV energy range and discuss
the implications for estimates of the intergalactic extinction due to pair 
production processes of the VHE photons on the Diffuse Extragalactic 
Background Radiation.
\end{abstract}
\keywords{ BL Lacertae objects: individual:
Mkn 501 -- Gamma rays: observations -- X-rays: galaxies}
\sloppy
\section{Introduction}
The BL Lac object Mkn 501 ($z=$0.034) 
underwent a major outburst in X-ray and in the VHE bands during 1997. 
Observations with the X-ray telescopes on board the BeppoSAX and
the RXTE satellites showed, compared to earlier data, 
a flux increase at several keV by up to one order of 
magnitude and a Spectral Energy Distribution
(SED) peaking frequently at very high energies, 
namely in the energy range above $\sim$25~keV
(Pian et al.\ \cite{Pian:98};  Lamer \& Wagner \cite{Lame:98}). 
In the VHE band Mkn 501 was the brightest known source in the
sky showing dramatic flux variability from a fraction to approximately
$\sim$10 times the flux of the Crab Nebula
(Aharonian et al.\ \cite{Ahar:99a}, called A99a in the following;
Aharonian et al.\ \cite{Ahar:99c}; 
Bhat et al.\ \cite{Bhat:97}; Catanese et al.\ \cite{Cata:97}; 
Djannati-Atai et al.\ \cite{Djan:99};
Hayashida et al.\ \cite{Haya:98}), with photon energies up to $\sim$20~TeV 
(Aharonian et al.\ \cite{Ahar:99b}, called A99b in the following).

The high energy nonthermal continuum emission of BL Lac objects
is widely believed to originate in a relativistic jet due to a population of high
energy electrons, emitting synchrotron radiation at longer wavelengths and
higher energy photons in Inverse Compton (IC) processes of the highest energy
electrons with lower energy seed photons (see for recent reviews 
Coppi \cite{Copp:97}; Sikora \cite{Siko:97}; Ulrich et al.\ \cite{umu:97}). The
origin of the IC seed photons has not yet been established. 
In so-called ``Synchrotron Self Compton'' (SSC) models the target photon 
population is dominated by low energy synchrotron photons 
(e.g.\ Bloom \& Marscher \cite{Bloo:93}; Ghisellini et al.\ \cite{GhisMD:1996}; 
Mastichiadis \& Kirk \cite{Mast:97}). In ``External Compton'' models the seed photons 
originate outside the emission volume and are, e.g.\ radiation from the 
nuclear continuum scattered or reprocessed in the broad-line regions 
(see e.g.\ Sikora et al.\ \cite{Siko:94})
or accretion disc photons (Dermer \& Schlickeiser \cite{Derm:94}). 
Simultaneous observations of the highest energy synchrotron photons in the X-ray 
band and of the highest energy IC photons in the VHE band make it possible 
to infer complementary information about the rapidly evolving population of highest 
energy electrons. Besides studies of large populations of similar sources, 
only detailed studies of the temporal and spectral
characteristics of individual sources over a broad wavelength region promise to
yield sufficient constraints to unambiguously identify the mechanism responsible for
the observed emission.
A further necessity to study in detail the temporal emission characteristics arises
from the fact that the observed VHE $\gamma$-ray spectra are expected to be substantially
modified by the intergalactic extinction due to pair production processes of the VHE
photons with the Diffuse Extragalactic Background Radiation (DEBRA) 
(Nikishov \cite{Niki:62}; Gould \& Schr\'{e}der \cite{Goul:65}; 
Stecker et al.\ \cite{Stec:92}).
The temporal analysis should yield enough redundant information not only to identify
the emission mechanism but also to determine the jet parameters making it possible
to infer the electron spectrum from its
X-ray synchrotron emission and to predict the intrinsic VHE spectrum.
The comparison of the intrinsic and the observed spectra 
yields the intergalactic extinction and as a consequence an estimate of the DEBRA density 
in the relatively unconstrained 0.5~$\mu$m to 50~$\mu$m wavelength region.
The observations with the HEGRA stereoscopic system of Cherenkov telescopes of 1997
showed that the Mkn 501 time-averaged VHE energy spectrum extends to energies 
well above 10~TeV and made it possible to sample an energy spectrum with an exponential cutoff
deeply into the exponential regime (A99b).
From 500~GeV to $\sim$20~TeV the spectrum can be described 
by a power law model with an exponential cutoff: 
$d{F}/d{E}\,\propto\,{(E/{\rm 1~TeV})}^{-0.9}\,\exp{(-E\rm /6.2\,TeV)}$.
In this paper we will study in detail the question whether the cutoff is caused 
by intergalactic extinction or by the emission mechanism itself.
Note that the HEGRA telescope system achieves an energy flux 
sensitivity $\nu\,f_\nu$ at 1~TeV of 10$^{-11}\,\rm erg/cm^2\,s$ 
for 1 hour of observation time. Furthermore, it is possible 
to determine differential spectra with reasonable statistical accuracy
for integration times of a few hours for sources with strong, Crab like VHE $\gamma$-ray flux levels.
Such Cherenkov telescope installations together with pointed X-ray telescopes 
like BeppoSAX, RXTE, or ASCA make it possible to study the temporal and spectral
properties of BL Lac objects on time scales of hours.

Earlier discussions of the implications of the 1997 Mkn 501 X-ray and VHE data can be 
found in Pian et al.\ (\cite{Pian:98}), Tavecchio et al.\ (\cite{Tave:98}),
A99a-b, Bednarek \& Protheroe (\cite{Bedn:99}), Hillas (\cite{Hill:99}), and
Konopelko et al.\ (\cite{Kono:99}). 
We present in this paper the analysis of a large data base of 
RXTE observations of Mkn 501 during 1997 and combine it with the
simultaneous and nearly simultaneous VHE data from A99a,b. 
Based on the spectral and temporal X-ray data presented 
in this paper and the spectral information measured with the HEGRA telescopes
we re-examine the constraints on SSC scenarios and their model 
parameters (as given e.g.\ in Tavecchio et al.\ \cite{Tave:98}; 
Bednarek \& Protheroe \cite{Bedn:99}).
Furthermore, following an approach described in Coppi \& Aharonian 
(\cite{Copp:99})
and Hillas (\cite{Hill:99}) we present SSC fits to the RXTE and BeppoSAX
data which yield, together with the HEGRA data, information on the
degree of intergalactic TeV Gamma-ray extinction.

The paper is structured as follows.
The data sample and the analysis of the RXTE and the 
HEGRA data is described in Sect.\ \ref{DATA}. 
The RXTE results and the correlations of the X-ray and VHE $\gamma$-ray
flux levels and spectra are investigated in Sect.\ \ref{RES}.
In Sect.\ \ref{INTER} we describe possible SSC model scenarios and 
SSC fits to the data. In Sect.\ \ref{DISC} we summarize the results. 
\section{RXTE and HEGRA observations and analysis} \label{DATA} 
We report on public RXTE observations 
which were performed from April 3rd,  1997 to July 14th, 1997.  
During April and May two observations were
made each night. The integration time of all pointings were between 10 and 70 
minutes per pointing. 
The RXTE satellite carries two pointed X-ray experiments (Bradt et al.\ \cite{Brad:93}), i.e.\ the
Proportional Counter Array (PCA) sensitive in the 2~keV to 100~keV energy range
with good sensitivity below 25 keV,  and the High Energy X-ray 
Timing Experiment (HEXTE) 
with sensitivity in the energy range from 15~keV to 150~keV.
Due to the very limited statistical information in the HEXTE data
and possible problems with the absolute HEXTE flux normalization and response matrices
we used only the 3-25~keV PCA data for the spectral fits.
After applying the standard screening criteria, the spectra were extracted
with {\sl FTOOLS 4.1} using bright source background models, and spectral fits were
performed with XSPEC 10.0. 
A constant neutral hydrogen column density of 2~$\rm \times~10^{20}~cm^{-2}$ was chosen,
a value which lies between the 21 cm line HI result 
of 1.73~$\rm \times~10^{20}~cm^{-2}$ (Stark et al.\ \cite{Star:92}) and the
ROSAT spectral absorption result of $\rm 2.87~\times~10^{20}~cm^{-2}$ (Lamer et al.\ \cite{Lame:96}).  
Since the analysis is restricted to the energy region above 3~keV the chosen
hydrogen column density has only a minor influence on the fitted spectra.
The majority of measurements are satisfactorily fitted with single power law models; 
for days with long integration times and high count rates we need broken power law
models to adequately describe the data. 

The HEGRA Cherenkov telescope system (Daum et al. \cite{Daum:97}; Konopelko et al.\ 
\cite{Kono:99a}) is located on the Roque de los Muchachos on the Canary Island of 
La Palma (lat.\ 28.8$^\circ$ N, long.\ 17.9$^\circ$ W, 2200 m a.s.l.). 
During monitoring of Mkn 501 from March 16th to October 2nd, 1997, 
110~h of high quality data was taken. The analysis tools and the estimate of the
systematic errors on the differential $\gamma$-ray energy spectra 
are discussed in A99a and A99b.
The results of fits to the differential spectra determined for 63 individual days are given
in A99a; the 1997 time-averaged spectrum over the energy range form 500~GeV to 24~TeV is given
in A99b.
\section{The Mkn 501 X-ray and VHE $\gamma$-ray characteristics during the 1997 outburst}
\label{RES}
\begin{figure}
\resizebox{\hsize}{!}{\includegraphics{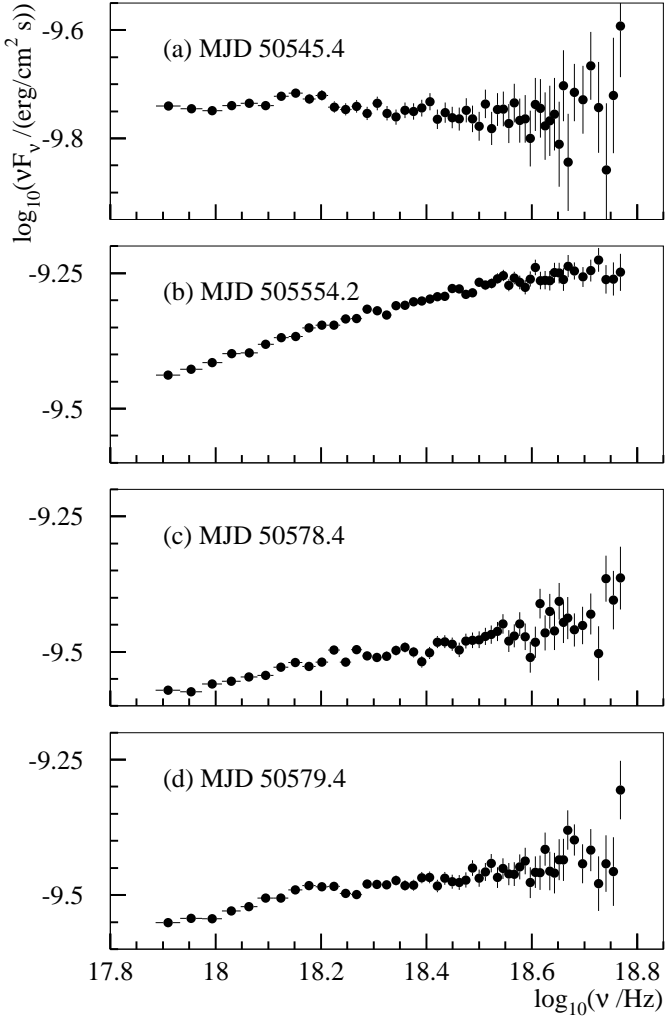}}
\caption
{Four RXTE 3-25~keV SEDs. The given MJD is centered on the observation window. From above to below the spectra
have been measured with 15~min, 18~min, 29~min, 
and 15~min integration time.}
\label{examples}
\end{figure} 
Fig.\ \ref{examples} shows four Mkn~501 SEDs as determined from the  
3-25~keV RXTE PCA data. It can be clearly recognized that falling SEDs with
spectral indices ($\alpha$ from $F_\nu\,\propto\,\nu^{-\alpha}$) 
larger than one, as well as increasing SEDs with spectral 
indices smaller than 1 have been observed. 
While the spectrum of MJD 50578 can be described satisfactorily by a pure power
law model, the spectra of MJD 50545, 50554, and 50579 show evidence 
for spectral softening with increasing energy.
For analyzing the trends in the spectral evolution we characterize
the 3-25~keV spectral steepness by a single power law index.
Note that during the 1997 outburst the position of the SED peak in the 
X-ray energy region varied substantially on a time scale of weeks: 
while for MJD 50545 (April 7th, Fig.\ \ref{examples}a) the peak is found at $\sim 5$~keV, 
the RXTE data as well as the BeppoSAX data of MJD
50554 (April 16th, Fig.\ \ref{examples}b) show a peak above 25\,keV. 
\begin{figure}
\resizebox{\hsize}{!}{\includegraphics{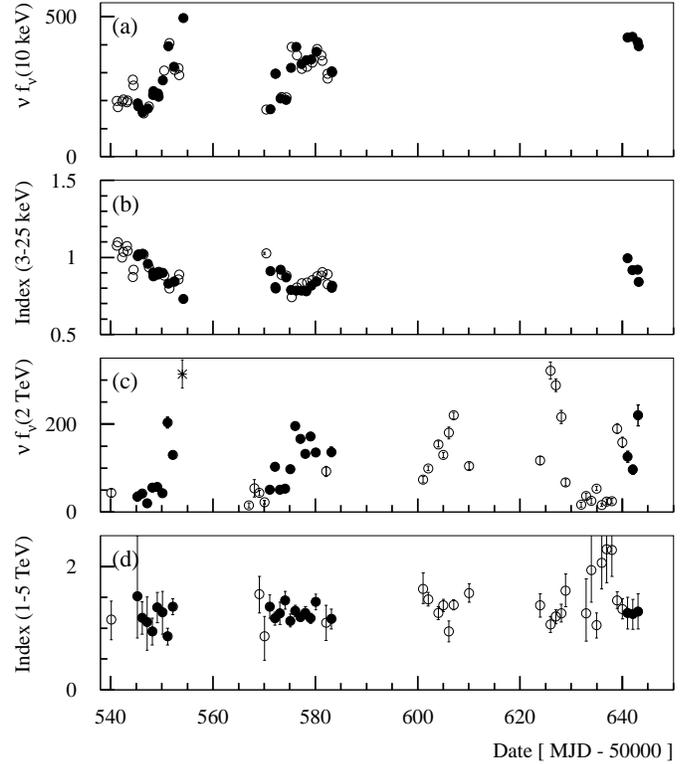}}
\caption
{The RXTE 10~keV energy flux levels in $\rm 10^{-12} erg/cm^2 s$ {\bf (a)}, 
the RXTE 3-25~keV spectral indices {\bf (b)},
the HEGRA flux levels at 2~TeV in $\rm 10^{-12} erg/cm^2 s$ {\bf (c)}, and
the HEGRA 1-5~TeV spectral indices {\bf (d)} as function of time.
For April 16th, 1997 (MJD = 50554) the VHE activity has been estimated 
from the CAT measurement (shown by an asterisk).
X-ray and VHE $\gamma$-ray observations taken
with less than 6~h time difference are highlighted by solid symbols.}
\label{lightcurve}
\end{figure} 

Fig.\ \ref{lightcurve}a-b show the fitted 
RXTE X-ray energy flux at 10~keV and the 3-25~keV 
spectral index as function of time.
The flux varies by a factor of three with 
shortest exponential increase/decay times\footnote{
For two flux levels $F_1,\,F_2$ measured at times $t_1$, $t_2$, the
exponential increase/decay time $\tau$ is defined as
$\tau\,\equiv\,(t_2\,-\,t_1)/\ln{(F_2\,/\,F_1)}$.}
of roughly one day.
The spectral index varies from 0.7 to 1.1 with typical velocities of 0.01/h.
We do not find any evidence for faster hardening than softening.

The HEGRA energy flux at 2~TeV and the 1-5~TeV spectral index $\alpha$
as function of time are shown in Fig.~\ref{lightcurve}c-d.
On April 16th, RXTE as well as BeppoSAX observed 
an extremely bright X-ray flare. Unfortunately, no HEGRA observations were
possible, and the flux at 2~TeV has here been estimated from lower energy 
observations with the CAT Cherenkov telescope (Djannati-Atai et al.\ \cite{Djan:99}).
Although the VHE Mkn 501 flux showed variability by factors of up to 
$\sim$~30 over the whole observation period, and by factors of up to 
10 for the days with nearly simultaneous RXTE
observations, the spectral index -- determined with an accuracy of typically 0.1
-0.3 -- is rather stable and statistically consistent with a constant value.  
Aharonian et al.\ (A99a, A99b) do not find any indication for a correlation 
of the absolute flux and the spectral shape.
From 1~TeV to 5~TeV the statistical accuracy of the observations 
is rather high, namely 
$\simeq$0.05 in the spectral indices of flux selected spectra.
Between 500~GeV and 1~TeV and from 10~TeV to 15~TeV 
the statistical accuracy is rather modest and changes of the spectral 
index in these energy regions 
by as much as 0.3 can not be excluded.
Remarkably, the mean 1-5~TeV spectral index of the data taken during phases of rapidly rising TeV flux 
differs by 0.10$\pm$0.06 from the mean index of the data taken during phases 
of rapidly falling TeV flux, weakly indicating a harder spectrum during phases of increasing
fluxes.
The TeV data shows shortest exponential increase/decay times of about 15~h; some evidence has been found for TeV variability on shorter time scales of a few hours
(A99a; Quinn et al.\ \cite{Quin:99}). 

On 25 days the time separation between the X-ray and the VHE $\gamma$-ray observations was smaller than 12~h 
and for 22 days it was smaller than 6~h. X-ray/VHE data pairs with time delays less 
than 6~h are marked in Fig.~\ref{lightcurve} by solid symbols.
Since the X-ray and the VHE variability is found on time scales slower \uga~1/2~day
and the spectral indices changed typically by less than 0.01 units per hour
these measurements are well suited for a meaningful correlation analysis.
Note that data with a clear signature of a substantially increasing or 
decreasing flux are of utmost importance for 
extracting information about the acceleration and emission mechanisms.
The X-ray/VHE observations during April, May, and June cover 
a considerable number ($\sim5$) of distinct strong flares including several phases
of substantial flux increase and decrease.

\begin{figure}
\resizebox{\hsize}{!}{\includegraphics{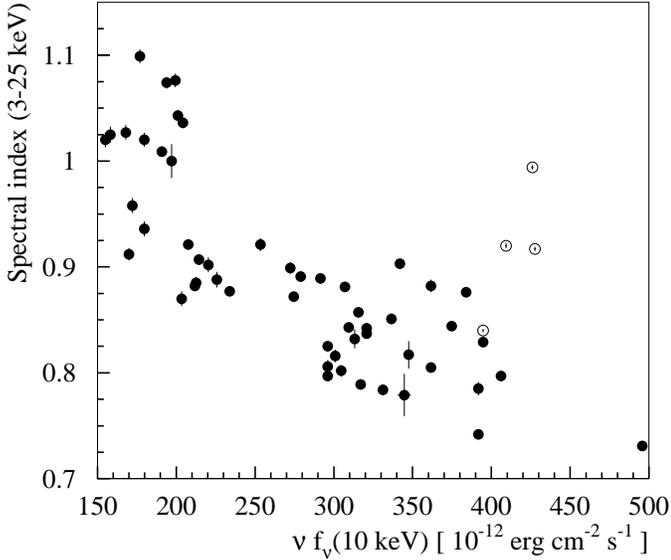}}
\caption
{The 3~keV to 25~keV spectral index against the 10~keV energy flux.
The April and May data is shown by solid symbols, the July data by open symbols.}
\label{xxcorr}
\end{figure} 
\begin{figure}
\resizebox{\hsize}{!}{\includegraphics{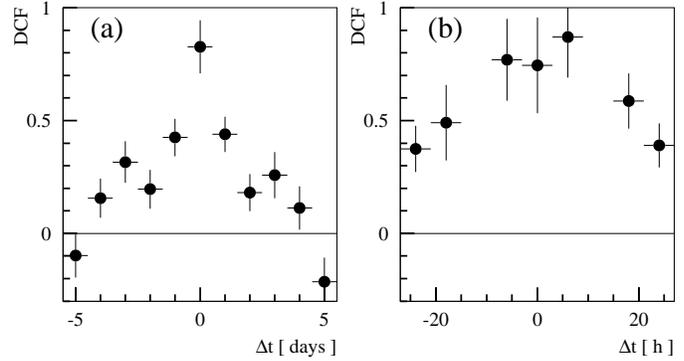}}
\caption
{The Discrete Correlation Function determined from the 3~keV and 25~keV RXTE fluxes
with one day bins {\bf (a)} and with 6~h bins {\bf (b)}.
The calculation of the flux values is based on the power law fits to the RXTE PCA data.
Positive $\Delta t$ correspond to the 3~keV flux variability leading the 25~keV flux
variability.}
\label{xxdcf}
\end{figure} 
\begin{figure}
\hspace*{1.5cm}
\resizebox{5.cm}{!}{\includegraphics{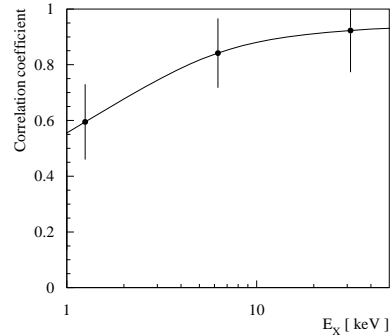}}
\caption
{The correlation coefficient of the X-ray flux at energy $E_{\mbox{\scriptsize X}}$ and the VHE $\gamma$-ray flux at 2~TeV
as function of $E_{\mbox{\scriptsize X}}$. The X-ray fluxes at $E_{\mbox{\scriptsize X}}$ have been determined from the
power law fits to the 3-25~keV RXTE data. 
Only data pairs with a time difference between X-ray and VHE observation
of less than 6 hours have been used.  }
\label{etcorr}
\end{figure} 
The X-ray data shows a tight correlation of the absolute flux and the
spectral index (Fig.~\ref{xxcorr}). An obvious tendency is that 
higher fluxes are accompanied by harder spectra. 
Compared to the April and May data, relatively high
X-ray fluxes during July were observed with -- in comparison to the correlation
shown by the April and May data -- rather soft spectral indices, indicating that
conditions in the source might have changed from April/May to July.
A possible time lag between the 25~keV flux and the 3~keV flux has been searched
for using the Discrete Correlation Function (DCF) (Edelson \& Krolik \cite{Edel:88}).
The analysis on a time scale of days shows that the
time lag is smaller than one day (Fig.~\ref{xxdcf}a). 
On a time scale of hours the DCF favors a time lag of the 
3~keV flux behind the 25~keV flux of smaller than $\simeq$15 hours
(Fig.~\ref{xxdcf}b).
Note that here and in the following 
the statistical error bars on the DCF have been computed
according to the prescription of Edelson \& Krolik (\cite{Edel:88})
from the deviations of the DCFs of individual data pairs from the mean DCF value
of the corresponding bin, taking thereby the quality of the correlation into account.
The errors on the DCF values resulting from the statistical errors of the measurements alone
are much smaller.

\begin{figure}
\resizebox{\hsize}{!}{\includegraphics{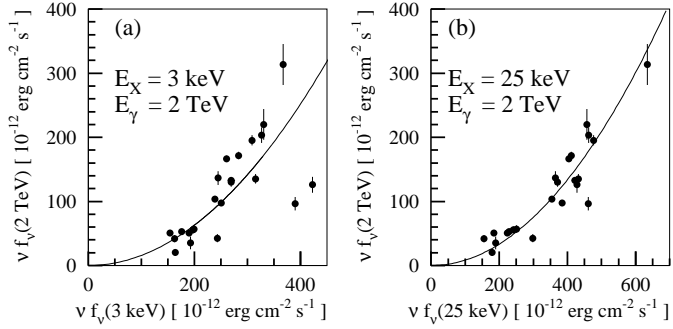}}\hfill
\caption
{Correlation of the RXTE energy fluxes at $E_X~=~$3~keV and $E_X~=~$25~keV, left and right side respectively,
with the 2~TeV energy fluxes. The X-ray fluxes have been determined from the power law fits to the 
3-25~keV RXTE PCA data. The lines shows fits according to 
\protect{$\nu F_\nu({\rm 2~TeV})\,\propto\,(\nu F_\nu({\rm E_X}))^2$}.
Only data pairs with a time difference between X-ray and VHE observation
of less than 6 hours have been used.}
\label{xtcorr}
\end{figure}

Fig.\ \ref{etcorr} shows the correlation coefficient of the X-ray 
flux at energy $E_X$ and the flux at 2~TeV as a function 
of $E_X$. Hereby the fluxes at $E_X$ were estimated from the 
power law fits to the RXTE data. 
For X-ray energies above several keV we find an excellent correlation with a 
correlation coefficient of about 0.8--0.9.
The statistical errors on the DCF do not allow us to decide whether
the correlation of the 3~keV and the 2~TeV or the 25~keV and the 2~TeV flux levels
is better.
Fig.\ \ref{xtcorr} shows the X/TeV-correlations for $E_X$ equal to 3~keV and 25~keV.
A quadratic relation between the keV and the TeV fluxes fits the data rather well.

\begin{figure}
\resizebox{\hsize}{!}{\includegraphics{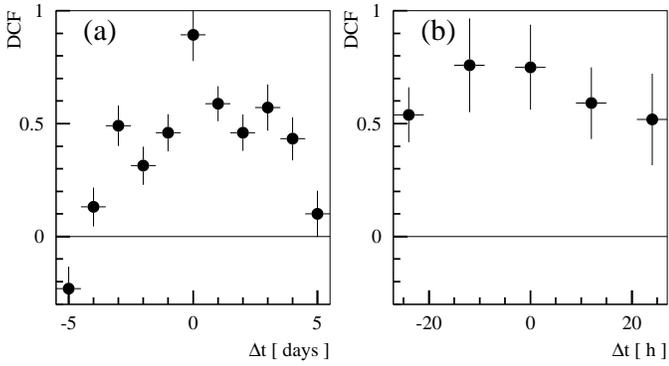}}
\caption
{The Discrete Correlation Function determined from the 25~keV RXTE and the 2~TeV HEGRA fluxes
with one day bins {\bf (a)} and 12~h bins {\bf (b)}.
Positive $\Delta t$ correspond to the 2~TeV flux variability leading the 
25~keV flux variability.}
\label{xtdcf}
\end{figure}

A DCF analysis of the 25~keV and 2~TeV fluxes shows no time lag on a timescale of days
(Fig.~\ref{xtdcf}a). 
An analysis on hour time scale indicates that the 25~keV variations happen
rather simultaneously with the TeV variations, or even lead them 
by several hours (Fig.~\ref{xtdcf}b). Although a time lag of the IC radiation 
relative to the synchrotron radiation is generally expected in SSC scenarios 
(Coppi \& Aharonian \cite{Copp:99}) the DCF of Fig.~\ref{xtdcf}b does not
yet allow definitive conclusions about the existence of such a time lag in Mkn~501.

One of the most interesting questions is whether the X-ray and VHE spectral indices
are correlated. The X-ray spectral index for all nights with VHE observations varies 
only by 0.25 which equals to a good approximation
the median accuracy of the HEGRA spectral index estimates. 
We draw the very important conclusion that VHE spectral variations 
with spectral changes comparable to the ones observed by RXTE are not 
excluded by the HEGRA data. The statistical accuracy of the HEGRA spectral estimates 
can be improved by determining average spectra for several days.
A closer look at the RXTE spectral indices however shows
that the mean spectral index of a sufficient large number of days varies by not more than 
$\simeq$0.1, a spectral difference which is again difficult to assess at TeV energies.
In our best try we grouped together HEGRA observations according to an 
RXTE spectral index below and above 0.85, 
with mean spectral indices which differ by 0.1. The HEGRA data gives a 1-5~TeV 
spectral index of 1.20$\pm$0.02$_{\rm stat}$ for the data sample with harder X-ray spectra,
and 1.28$\pm$0.04$_{\rm stat}$ for the data sample with softer X-ray spectra.
A difference of the VHE spectral indices of the same magnitude as at X-ray 
energies seems likely but the statistical evidence is marginal.
To summarize, the variations of the RXTE spectral indices are rather small 
compared to the accuracy of the HEGRA spectral estimates and the data 
supports rather than excludes a correlation of the X-ray/VHE spectral indices.
This is consistent with recent {claim} of the CAT group who reported 
that the joint spectrum of the data of the two strongest April flares, 
was harder than the 1997 mean spectrum (Djannati-Atai et al.\ \cite{Djan:99}).
\section{Interpretation in the framework of SSC scenarios}
\label{INTER}
In this section we first discuss possible SSC scenarios for explaining the 
1997 Mkn~501 X-ray and VHE characteristics and discuss the constraints 
on the model parameters (Sect.\ \ref{sscmodel}). 
Eventually we show several examples of fits to multiwavelength data and compare
the estimated emitted VHE spectrum with the observed one (Sect.\ \ref{debramodel})
to infer information about the intergalactic extinction caused by the DEBRA 
(Sect.\ \ref{debraresult}).
Note that the quadratic dependence of the VHE (probably IC) $\gamma$-ray fluxes on 
the X-ray (probably synchrotron) fluxes shown in Fig.~\ref{xtcorr}
favors a SSC scenario in which the production of seed photons is 
closely connected to the production of the nonthermal electron population.
\subsection{Constraints on the model parameters}
\label{sscmodel}
In the following a ``one zone model'' is used for simplicity;
the synchrotron and IC radiations originate in a spherical emission 
volume homogeneously filled with an isotropic population of relativistic electrons 
which move in a magnetic field characterized by its mean value $B$.

In the first scenario (``scenario 1''), the emission volume is characterized by a constant
Doppler factor ($\delta_{\rm j}^{-1}\,=\,\Gamma(1-\beta\,\cos{(\theta)})$,
with $\Gamma$ and $\beta$ the bulk Lorentz factors of the emitting volume 
and $\theta$ the angle to the observer), a rather constant magnetic field $B$, and
a slightly variable radius. While the spectrum of accelerated electrons
does not change with time, only the variability of the rate of accelerated particles
entering the emission region causes the observed X-ray and VHE $\gamma$-ray variability.
The X-ray synchrotron spectrum hardens when a large amount of freshly
accelerated particles enters the emission region (i.e. during the rising stage of a flare)
and subsequently steepens due to the cooling of these particles 
(i.e.\ during the decaying stage of a flare).
A continuously changing X-ray spectral index is expected in this scenario if
the cooling time of electrons responsible for the X-ray radiation is 
comparable to the flux variability time scale.
We envisage that the electrons responsible for the low energy X-ray emission 
(below 1\,keV or so) have no time to cool before they escape the emission region
and as a consequence the spectra are always very hard in this energy region. By contrast, 
at energies above $\sim25$~keV most often soft spectra are observed due to the
rapid cooling of the responsible electrons. 
Diffusive shock acceleration predicts a power law spectrum of nonthermal particles 
$dN_{\rm e}/dE~\propto~E^{-p}$ with $p\,\simeq\,2$ which 
is steepened at higher energies by one unit ($p'\,\simeq\,3$) due to synchrotron
cooling. Therefore, the produced synchrotron radiation has a spectrum with 
$F_\nu~\propto~\nu^{-\alpha}$ with $\alpha\,=\,(p-1)/2\,\simeq\,0.5$ in the low 
energy region and with $\alpha'\,=\,(p'-1)/2\,\simeq\,1$ in the high energy region.
Indeed, the Mkn 501 spectra observed with BeppoSAX during April 1997 show for all three 
days a spectrum $F_\nu$ with an spectral index $\alpha=0.5$ 
below several keV, and $\alpha'\approx 1$ 
above $\simeq\,$25~keV.
Furthermore, observations with OSSE taken between April 9th to April 15th, 1997
showed a hard 50~keV to $\sim$470~keV spectrum characterized by a spectral 
index $\alpha'$ of 1.1 (Catanese et al.\ \cite{Cata:97}).
In this scenario the X-ray spectral index is determined by the 
temporal evolution of the density of emitting particles rather 
than by the density itself. As a consequence, the correlation 
of the X-ray and VHE spectral indices should
be tighter than the correlation of the absolute VHE fluxes and VHE spectral indices, 
in accord with the results shown in the previous section.
Assuming an escape time of the low energy electrons smaller than several days, the
IC seed photon density approximately follows the rate of injected particles
and a faster than linear (but not necessarily quadratic) rise of the VHE fluxes 
for increasing X-ray fluxes is a natural
consequence of the changing injection rate of accelerated particles.

The following alternative scenario (``scenario 2'') seems attractive since
it automatically accounts for the $<1$keV X-ray spectrum with a spectral index $\alpha \approx 0.5$,
repeatedly observed with BeppoSAX (Pian et al.\ 1998a,b).
The basic difference to scenario 1 is a minimum energy $E_1$ of accelerated electrons 
which is responsible for the break of the synchrotron spectrum.
A large magnetic field leads to rapid cooling and hence
steepening of initial electron spectra down to the minimum energy $E_1$.
Below $E_1$ electrons which underwent cooling form a spectrum with the canonical
spectral index $p''=2$, independent of the value $p$ of the initial spectrum.
Above $E_1$ the electron spectrum is characterized by a spectral index $p'=p+1\approx 3$ for
$p\approx 2$.
In this scenario one obtains naturally a synchrotron spectrum with $\alpha=(p''-1)/2=0.5$ from
the electrons with energies below $E_1$.
Furthermore, the scenario allows for a high magnetic field which assures 
fast cooling (relative to the flux variability time scale)
of the electrons near the peak of the SED and thus a high radiation efficiency.
Spectral changes in the X-ray energy range are caused in this scenario
by the parameters $E_1$ and $p$ varying with time.

We explored in detail a third scenario (``scenario 3'') in which the observed characteristics 
are caused by a time-independent electron spectrum, but where all 3 parameters ($B$,$R$, and
the density of emitting particles) vary substantially.
A changing magnetic field together with a rather stable 
(e.g.\ due to cooling on small time scales) 
electron spectrum could be the reason for large spectral changes in the X-ray band 
accompanied by a stable VHE spectrum. Note, however, that it is necessary in this scenario
to introduce fixed relationships between the parameters characterizing the emission volume (i.e.\ 
between $R$ and $B$) in order to explain the very tight X-ray/VHE correlation.
Further studies in this direction are underway.

Note that an alternative scenario in which the spectral variability as well as the 
X-ray and VHE intensity variations are caused by varying only the
relativistic Doppler factor $\delta_{\rm j}$ of the emission 
volume can be excluded from the X-ray data alone. 
As a consequence of the Lorentz invariance of $F_\nu/\nu^3$
a change of the energy at which the Mkn 501 SED peaks by a factor of $a$ 
would be accompanied by a change of the observed flux by 
$\approx\,a^{3+\alpha}$, where $\alpha$ is the spectral index 
at the frequency of observations ($F_\nu\,\propto\,\nu^{-\alpha}$).
The spectra of Fig.~\ref{examples} show a change of the SED peak position by approximately 
a factor of 5 but clearly without the corresponding increase 
in luminosity.

Deciding on one of the first three scenarios will be crucial for a final
understanding of the Mkn~501 SEDs.
In the following we will summarize the constraints on the model 
parameters. The significant VHE $\gamma$-ray 
variability on timescales $\Delta t_{\rm obs}$ 
of approximately half a day implies the well known upper limit on the radius of the emitting
volume:
\begin{equation}
R \,<\delta_{\rm j}\,c\,\Delta t_{\rm obs}\,=\,
1.3\cdot10^{16}~(\delta_{\rm j}/10)~(\Delta t_{\rm obs}/12 \textrm{h})~~
\textrm{cm}
\label{SIZE}
\end{equation}
Given this limit, 
the Doppler factor is constrained by the condition of a negligible or modest 
absorption of high energy photons in pair production processes with lower energy 
synchrotron photons inside the emission volume.
Extrapolating the synchrotron spectrum observed by BeppoSAX at soft X-ray energies (\uka 1~keV) 
according to $F_\nu\,\propto\,\nu^{-\alpha}$ with $\alpha=0.5$
towards lower frequencies, we obtain (see e.g.\ Svensson \cite{Sven:87})
an optical depth as function of the photon energy $E_{\gamma}$ (frame of observer) 
of the gamma-ray:
\begin{equation}
\tau_{\gamma \gamma}\,\simeq\,(\Delta t_{\rm obs}/{\rm 12~h})^{-1}\,
(E_{\rm \gamma}/{\rm 10~TeV})^{0.5}\,(\delta_{\rm j}/4.8)^{-5}~.
\label{PA}
\end{equation}
Here and in the following we assumed a Hubble constant of 60~km~s$^{-1}$~Mpc$^{-1}$.
If the (long wavelength) target photon density in the source changes as the observed 
X-ray and VHE luminosity, one would expect a steepening VHE spectrum for large VHE luminosities. 
The non-observation of such
a correlation with an accuracy of better than 0.1 in the VHE spectral index limits the 
optical depth at 10~TeV to values below $\sim$1/4, resulting in a lower limit on the 
Doppler factor of $\delta_{\rm j}$~\uga~6.3.

The X-ray and VHE $\gamma$-ray flux levels (Fig.~\ref{xtcorr}) 
indicate that the electron cooling is dominated by synchrotron 
cooling rather than by IC cooling.
For the case that the spectral evolution after flares is dominated by 
synchrotron cooling, Takahashi et al.\ (\cite{Taka:96}) 
and Kirk et al.\ (\cite{Kirk:98}) 
discussed the possibility to constrain the magnetic field from observations of
time lags between high and low energy synchrotron radiation.
Due to the energy dependent cooling times of electrons with energy $E_{\rm e}$:
$t_{\mbox{\scriptsize S}} \, = \,\left[(4/3)~\sigma_\textrm{\small T}~c~(B^2/8\pi)~E_{\rm e}/
m_{\rm e}^2\,c^4\right]^{-1}$ with $\sigma_\textrm{\small T}$ the Thomson cross section,
one expects that after a flare the lower energy synchrotron radiation 
(from electrons of energy $E_{\rm l}$) lags the higher energy synchrotron 
radiation (from electrons of energy $E_{\rm h}$) by
$\Delta t_{\rm s} = t_{\rm s}(E_{\rm l}) - t_{\rm s}(E_{\rm h})$.
As a consequence, the DCF of the lower and higher energy radiation should show a shoulder
reflecting this time lag.
Using the delta functional approach for the energy of synchrotron photons 
$E_{\mbox{\scriptsize S}}$ produced by electrons of energy $E_{\rm e}$:
\begin{equation}
E_{\mbox{\scriptsize S}}\,\simeq~200~(\delta_{\rm j}/10)~(B/\textrm{1~G})~
(E_{\rm e}/\textrm{1~TeV})^2~\rm keV,
\label{SE}
\end{equation}
and a shoulder of the DCF of the 3~keV and the 25~keV radiation up to 
$\Delta t_{\mbox{\scriptsize S}}\,\approx\,15$~h (Fig.~\ref{xxdcf})
yields:
\begin{equation}
B\,\simeq\,0.025\,(\Delta t_{\mbox{\scriptsize S}}/\textrm{15~h})^{-2/3}\,
(\delta_{\rm j}/10)^{-1/3}~G.
\label{BV}
\end{equation}
Since several other reasons are expected to broaden the DCF
(e.g.\ delays due to the size of the emission region)
this value of the magnetic field should be regarded as a rough lower limit.

In the reminder of this subsection we discuss which X-ray and VHE photons are 
produced by the same electrons and are henceforth expected to show correlated 
intensity variations.
We find that the IC photons near and above the energy at which the gamma-ray SED 
peaks are mainly produced in Thomson scatterings as well as in the 
Thomson to Klein-Nishina (KN) transition regime. 
This can qualitatively be understood as follows 
(a more quantitative discussion is given in the next subsection, 
see also Zdziarski \cite{Zdzi:89}).
On the one hand, due to a probable seed photon energy flux rising approximately as 
$\nu\,F_\nu\,\propto\,\nu^{0.5}$ the Thomson processes yield 
the largest energy flux of IC photons for the highest 
seed photon frequencies. On the other hand, due to a seed photon number 
spectrum falling steeply like $d$N/$d\nu\,\propto\,\nu^{-1.5}$ 
with increasing frequency, and due to the suppression of the KN 
scatterings inversely proportional to the centre of momentum energy of seed photon
and electron the KN production of VHE photons is strongly suppressed.
Electrons of energy $E_{\rm e}$ produce therefore the largest
IC power per logarithmic bandwidth interacting with photons from the energy 
(all energies in the frame of the emitting volume)
\begin{equation}
\varepsilon_0\,\approx\,1/4\,(m_{\rm e}\,c^2)^2/E_{\rm e}\,=\,1/16\,\cdot (E_{\rm e}/{\rm 1~TeV})^{-1}~\rm eV
\label{E0}
\end{equation}
producing Gamma-rays of energy (frame of observer):
\begin{equation}
E_{\mbox{\scriptsize IC}}\,\approx\,\delta_{\rm j}\,\eta\,E_{\rm e}~.
\label{IE}
\end{equation}
Obviously, for IC scattering in the extreme KN regime it is $\eta\,\approx\,1$; 
for Thomson scattering on photons of energy $\varepsilon_0$
one naively expects from the delta functional approximation 
of the IC cross section (e.g.\ Ginzburg \& Syrovatski \cite{Ginz:64})
a value $\eta$ of 1/3. 
From our simulations we find that electrons of energy $E_{\rm e}$ indeed 
produce the maximum energy per logarithmic bandwidth 
at IC photon-energies given by Eq.\ (\ref{IE}) 
with $\eta \approx 1/3$ near the peak of the SED.
Note that in our models the synchrotron self absorption cutoff is typically found 
between 10$^{11}$~Hz and 10$^{12}$~Hz (laboratory frame), 
well below the frequencies of the most important seed photons.

Eqs. (\ref{SE}) and (\ref{IE}) can be combined to derive the relation of the
observed synchrotron and IC photon energies 
produced by electrons of the same energy region:
\begin{equation}
E_\textrm{\scriptsize IC}/{\rm TeV}\,\approx\,
\left(\frac{\delta_{\rm j}/10}{B/\textrm{0.05~G}}\right)^{1/2}\,\,
(E_\textrm{S}/\textrm{1~keV})^{1/2}.
\label{mapping}
\end{equation}
It can be recognized that 
the ``mapping'' between synchrotron and IC photon energies is only a weak function
of $\delta$ and $B$. 
It should be noted that Eqs.\ (\ref{SE}), (\ref{IE}), and (\ref{mapping}) are only rough
approximations; in reality, electrons of a certain energy produce synchrotron and IC 
photons with energies scattered over more than a magnitude.
\subsection{SSC fits to multiwavelength data}
\label{debramodel}
\begin{figure}
\resizebox{\hsize}{!}{\includegraphics{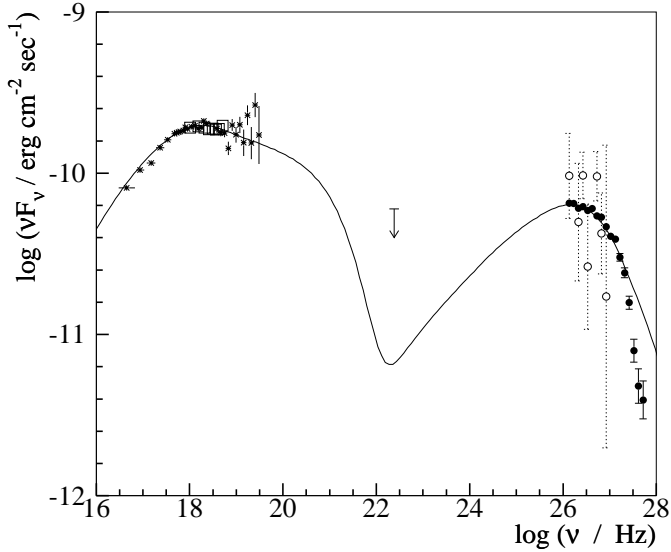}}
\caption
{BeppoSAX (asterisks), RXTE (squares), and HEGRA (open circles) 
observations of April 7th, 1997 (MJD 50545)
and the HEGRA 1997 time-averaged Mkn~501 spectrum scaled according to the detection rate of this day 
(full circles). 
The line shows a SSC model for 
a Doppler factor $\delta_{\rm j}\,=\,25$, a magnetic field $B\,=\,0.037$,
and a radius of $R\,=\,1.5 \cdot 10^{16}$\,cm. 
On April 7th, 1997, the HEGRA observations were performed from UTC 4:53 to 5:27.
Two RXTE observations were made, the first from UTC 7:16 to 7:31 and the second from 
UTC 10:37 to 10:52 showing to an accuracy of 0.008 no significant change of the 
X-ray spectral index.
The BeppoSAX observations were performed from UTC 5:30 to 16:00.
The BeppoSAX LECS, MECS,  and PCA data has been normalized to fit 
the RXTE data at energies around 5 keV.
Compared to the spectrum shown in Pian et al.\ (\cite{Pian:98}) 
we reduce the PCA normalization by 
35\% which eliminates the apparent 
discontinuity of the BeppoSAX spectra at $\simeq$15~keV and is then consistent with the 
spectral shape measured from 3~keV to 25~keV with RXTE.
The 2$\sigma$ upper limit at 2.4$\cdot 10^{22}$~Hz has been derived from
EGRET observations between April 9th and April 15th, 1997 under the assumption of a constant
emission level (Catanese et al.\ \cite{Cata:97}).}
\label{scen1a}
\end{figure}

\begin{figure}
\resizebox{\hsize}{!}{\includegraphics{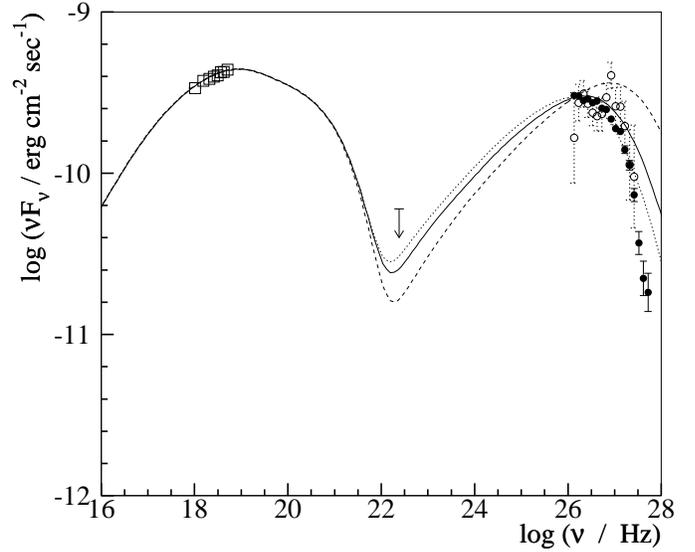}}
\caption
{RXTE (squares), and HEGRA (open circles) observations of April 13th, 1997 (MJD 50551)
and the HEGRA 1997 time-averaged Mkn~501 spectrum scaled according to the detection rate of this day 
(full circles). The solid line shows a SSC model for  
a Doppler factor $\delta_{\rm j}\,=\,25$, 
a magnetic field $B\,=\,0.037$, and a radius of $R\,=\,1.1 \cdot 10^{16}$\,cm. 
The dashed and dotted lines show the uncertainty in the shape of the
predicted VHE spectrum arising from changing the value $\sqrt{\delta_{\rm j}/B}$ 
(dashed line: $\delta_{\rm j}=100$, $B=0.012$~G,
dotted line: $\delta_{\rm j}=25$, $B=0.12$~G).
On April 13th, 1997, the HEGRA observations were performed from UTC 1:37 to 2:55.
Two RXTE observations were made, the first from UTC 6:24 to 6:53 and the second from 
UTC 11:01 to 11:54 showing to an accuracy of 0.006 no significant change of the 
X-ray spectral index.
The 2$\sigma$ EGRET upper limit at 2.4$\cdot 10^{22}$~Hz 
is the same as in Fig.~\ref{scen1a}.}
\label{scen1b}
\end{figure}

Figs.\ \ref{scen1a} and \ref{scen1b} show simultaneous and nearly simultaneous 
RXTE, BeppoSAX, and HEGRA data taken on April 7th and April 13th, 1997 respectively
together with the results from the SSC models.
Note that the RXTE spectrum from April 7th is with a spectral index of 
1.1 among the softest spectra observed with RXTE in 1997 while the spectrum of 
April 13th has a rather average spectral index of 0.83.

The SSC model is computed with the time-independent part of the 
code described by Coppi (\cite{Copp:92}). 
This time-independent part is similar to the codes described e.g.\ by 
Inoue \& Takahara (\cite{Inou:96}) and Kataoka et al.\ (\cite{Kata:99}).
For a given set of parameters ($\delta_{\rm j}$, $B$, $R$) the intrinsic VHE spectrum can be
computed by deriving the spectrum of the emitting electron population from the synchrotron spectrum 
(Coppi \& Aharonian \cite{Copp:99}). 
This ``inversion'' of the synchrotron spectrum yields unique results for 
sufficiently smooth electron spectra. 
Unfortunately, the predicted VHE spectrum depends partially on the electron spectrum outside the
energy region constrained by the X-ray observations. Given the observational evidence and
the theoretical framework outlined above, 
we use the following probable assumptions about the 
synchrotron spectrum outside the energy region covered by the RXTE observations, 
namely below 0.1~keV and above 25~keV.
The RXTE observations, 
as well as the OSSE observations from April 9th to April 15th, 1997
show spectra with spectral indices of \uka1.1.
We choose accordingly a spectrum of accelerated electrons with
$dN_{\rm e}/dE~\propto~E^{-p}$ with $p~=~2.2$.
Electrons which did not have time to cool produce a synchrotron spectrum at energies below 
0.1~keV with $F_\nu\,\propto\,\nu^{-0.6}$. Above 25~keV the cooled electrons 
give a spectrum with $F_\nu\,\propto\,\nu^{-1.1}$. 
First we assume a Doppler factor $\delta_{\rm j}\,=\,25$ 
and a magnetic field $B\,=\,0.037$~G which lie well above the lower limits from
Eqs.\ (\ref{PA}) and (\ref{BV}); 
later we show how results change varying these parameters in a wide margin.
For this parameter values the 3-25~keV radiation
is produced by roughly the same electrons as the 3-9~TeV emission 
(see Eq.\ \ref{mapping}) in accord with the tight X-ray/VHE correlation shown in 
Figs.\ \ref{etcorr} and \ref{xtcorr}.

The synchrotron spectrum probably cuts off at some energy. 
The BeppoSAX and the OSSE observations do not show any indications for such
a cutoff up to energies of at least 150~keV. Importantly, for the used values of
$\delta$ and $B$ the location of this cut off does not influence the SSC prediction 
of the IC spectrum up to energies of 25~TeV 
as long as the cut off does not influence the synchrotron spectrum below 
$\sim$150~keV (compare Eq.\ (\ref{mapping})).
So in the following we use a rather high exponential cutoff at 10~MeV.
Note that 10~MeV synchrotron photons can be produced by $\sim$20~TeV 
electrons (Eq.\ (\ref{SE})) which is still below the maximum energy
of electrons expected for the case of shock acceleration with Bohm diffusion
(e.g.\ Hillas \cite{Hill:99}).

The predicted IC to synchrotron luminosity ratio depends on the assumed 
radius of the emission volume as 1/$R^2$. 
Assuming negligible intergalactic extinction at 500~GeV we determine a radius 
of $1.5 \cdot 10^{16}$\,cm and $1.1 \cdot 10^{16}$\,cm 
for the April 7th and April 13th data, respectively.
Doppler factors below $\simeq$15 result in too large radii which 
conflict with the upper limit from the flux variability time scale (Eq.\ (\ref{SIZE})). 
Note however that actually there could be some intergalactic extinction at 500~GeV; one should keep
in mind that the fit predicts the shape rather than the absolute normalization of the intrinsic 
VHE spectrum.

Finally we turn to the SSC model spectra given in Fig.~\ref{scen1a} and 
Fig.~\ref{scen1b} (solid lines). 
The model calculations show that the break of the electron spectrum 
observed in the synchrotron radiation
at X-ray energies is expected to give rise to a 
significant curvature of the IC spectrum 
in the TeV energy region.
Obviously, the scenario favors a very small or alternatively an energy-independent 
intergalactic extinction of the radiation up to approximately 5~TeV.  
The observed time-averaged Mkn 501 spectrum shows additional steepening above 
5~TeV which is not found in the predicted spectrum. Possibly, this additional 
steepening is caused by the intergalactic extinction. 

The BeppoSAX, OSSE, and RXTE observations show that the X-ray spectrum turns over from a
spectral index $\alpha \approx 0.6$ below several keV to $\alpha' \approx 1.1$ above several
keV. As can be seen in Figs.\ \ref{scen1a} and \ref{scen1b} the 
corresponding electron spectrum produces an IC component 
with a power law behavior well below 500~GeV (spectral index $\alpha_{\mbox{\scriptsize IC}}=0.6$)
and well above several TeV (spectral index $\alpha'_{\mbox{\scriptsize IC}}=1.6$).
The power law shape of a low and high energy IC component in an energy region where the 
KN effect is not negligible was already discussed by Tavecchio et al.\ (\cite{Tave:98}).
The appearance of the IC power law components can be understood by noting that
the relative contribution to the production of IC photons from scatterings in the 
Thomson regime and from scatterings in the KN regime does not depend on the  
energy of the responsible electrons as long as the seed photon and electron 
populations are described by power law spectra in the relevant energy ranges\footnote{The 
statement can easily be derived from the properties of the
KN cross section, namely that the cross section as well as the fraction of 
electron-energy transferred to the IC photons only depends on the 
centre of momentum energy of seed photon and electron,
i.e.\ on the product of the seed 
photon and electron energy in the comoving frame of the
emission volume.
}.
As a consequence, 
the slope of the produced IC component can be derived by considering 
scattering processes in the Thomson regime alone and using the 
$\delta$-functional approximation of the IC cross section (see also Tavecchio et al.\ \cite{Tave:98}).
The rate of produced IC photons is given by
\begin{equation}
\dot{n}_{\mbox{\scriptsize IC}}(E_{\mbox{\scriptsize IC}})\,=\,c\!
\int_{E_1}^{E_2}\!\!\!
\int \! n_{\rm e}(E_{\rm e})\,n_{\rm ph}(\varepsilon)\sigma(E_{\rm e},\varepsilon,E_{\rm
\mbox{\scriptsize IC}})\,d\varepsilon\,dE_{\rm e},
\end{equation}
where $n_{\mbox{\scriptsize IC}}$, $n_{\rm e}$, and $n_{\rm ph}$ are the numbers of 
IC photons, high energy electrons, and seed photons per unit volume and per energy interval.
Using $dn_{\rm e}/dE_{e}  \,\propto \, E_{\rm e}\,\!^{-p}$ for $E_{\rm e}$ 
between the energies $E_1$ and $E_2$,
$dn_{\rm ph}/d\varepsilon \, \propto \, \varepsilon^{-(\alpha+1)}$ 
over the whole relevant energy region, 
and the rough approximation for the cross section for the production of IC photons
$\sigma=\sigma_{\mbox{\scriptsize T}}\,\delta\left[ E_{\mbox{\scriptsize IC}}-\frac{4}{3}\,\varepsilon
\left(\frac{E_{\rm e}}{m_{\rm e} c^2}\right)^2\right]$ we get
\begin{equation}
\dot{n}_{\mbox{\scriptsize IC}}(E_{\mbox{\scriptsize IC}})\,\propto \,
E_{\mbox{\scriptsize IC}}\,\!^{-(\alpha+1)}
\int_{\xi\,E_{\mbox{\scriptsize IC}}}^{E_2}\,\,E_{e}\,\!^{-(p-2\alpha)}\,\,dE_{\rm e}.
\label{APP}
\end{equation}
The lower bound of the integral assures with $\xi\approx 3$ scattering in the Thomson regime;
the additional contribution of IC photons produced in scatterings in the KN (transition) regime 
has formally been absorbed into a multiplicative correction factor, which is 
(as discussed above) to good approximation constant as long as 
$\xi\,E_{\mbox{\scriptsize IC}}\ll E_2$.
Although Eq.\ (\ref{APP}) is a crude approximation, we find that it
predicts the spectral slopes satisfactorily for a wide range of $p$ and $\alpha$ 
values provided $\alpha>0$ (see Zdziarski \cite{Zdzi:89} who
discusses in detail the applicability of the $\delta$-functional approximation).
For an electron population with spectral index $p$ 
interacting with its own synchrotron radiation, i.e.\ $\alpha=(p-1)/2$, one obtains the
well known result 
\begin{equation}
\dot{n}_{\mbox{\scriptsize IC}}(E_{\mbox{\scriptsize IC}})\,\propto\,
E_{\mbox{\scriptsize IC}}\,\!^{-(\alpha+1)} \ln{(E_2/(\xi E_{\mbox{
\scriptsize IC}}))},
\label{R1}
\end{equation}
namely that the IC spectrum has approximately the same spectral index 
$\alpha_{\mbox{\scriptsize IC}}$ as the 
synchrotron spectrum. A cooled electron spectrum with $p'\,\approx\,3.2$ which interacts with
the seed photon spectrum of the uncooled electrons
with $\alpha\,\approx\,0.6$ yields 
\begin{equation}
\dot{n}_{\mbox{\scriptsize IC}}(E_{\mbox{\scriptsize IC}})\,\propto\,
E_{\mbox{\scriptsize IC}}\,\!^{-(\alpha'_{\mbox{\tiny IC}}+1)}
\label{R2}
\end{equation} 
with $\alpha'_{\mbox{\scriptsize IC}}\,=\,p'-\alpha-1\,\approx\,1.6$.
Remarkably, Eqs.\ (\ref{R1}) and (\ref{R2}) show that  
in the case of SSC scenarios of TeV blazars
changes of the electron spectral index are reflected 
in an approximate 
one to one relationship in changes of the spectral index of the 
produced IC component.

Based on qualitative arguments about the stability of the Mkn~501 VHE
spectrum Konopelko et al.\ (\cite{Kono:99}) inferred an intrinsic TeV 
spectrum following a pure power law with a spectral index $\alpha'$ near one. 
Given the spectral indices $p=2.2$ and $p'=3.2$ of the low and the
high energy electrons respectively, as deduced from the X-ray spectra, 
Eq.\ (\ref{APP}) as well as our simulations show that it is 
extremely difficult to obtain such an IC spectrum in a SSC scenario 
since the resulting spectrum is characterized by $\alpha_{\mbox{\scriptsize IC}}=0.6$
or $\alpha'_{\mbox{\scriptsize IC}}=1.6$, i.e.\ it is too soft or too hard by $\sim$0.5 
units in the spectral index. 
Furthermore, in the scenarios discussed here, the narrow frequency range 
where the IC spectrum turns over ($\alpha\approx 1$) is a region where we expect more 
rather than less spectral variability than below and above this frequency range.

Since (i) the RXTE spectral index of April 13th approximately 
represents the mean value during the 1997 RXTE observations of this source and
(ii) the HEGRA VHE spectra did not vary during 1997 within statistical errors
we think it is justified to compare the prediction of VHE spectrum 
of this day with the observed 1997 time averaged VHE spectrum.
The predicted VHE spectrum mainly depends on the parameter 
$(\delta_{\rm j}/B)^{1/2}$ determining the ratio of the IC SED peak frequency 
and the square root of the synchrotron SED peak frequency 
(Eq.\ (\ref{mapping})).
The Doppler factor as well as the magnetic field are poorly constrained towards higher values. 
These large uncertainties however have only a limited impact on the model fits.
Combinations of large Doppler factors and small magnetic fields 
are constrained by the ``mapping relation'' between X-ray and VHE photon 
energies given in Eq.\ (\ref{mapping}).
Using an extremely high Doppler factor of $\delta=100$ together with the limiting
lower value of the magnetic field from Eq.\ (\ref{BV}) of 0.012~G results in
a mapping of the 1~TeV to 5~TeV radiation with the 0.02~keV to 0.6~keV radiation,
which seems unlikely, given the BeppoSAX, HEGRA, and CAT observations 
during April 1997 which showed dramatic variability above several keV and at
TeV energies but almost no variability at all below 1~keV.
The dashed line in Fig.~\ref{scen1b} shows the predicted VHE spectrum
for this limiting upper value of the parameter $(\delta/B)^{1/2}$.
The difference between the predicted and the observed $\gamma$-ray spectrum can be 
interpreted as an {\it upper limit} on the increase of intergalactic extinction with
$\gamma$-ray energy.
Additionally, the magnetic field is not constrained towards higher values.
The dotted line in Fig.~\ref{scen1b} shows 
the effect of increasing the magnetic field from 0.037~G 
to a limiting value of 0.12~G 
(which has the same effect on the shape of the predicted 
IC spectrum as decreasing the Doppler factor to a value of $\delta\simeq8$).
Larger magnetic field values (more precisely, smaller values of
$(\delta_{\rm j}/B)^{1/2}$) are highly improbable since the
resulting intrinsic VHE spectra are steeper than the observed ones
in the energy range above $\sim$1~TeV. Intergalactic extinction which
decreases for increasing energy is rather unlikely given our present 
understanding and knowledge of the DEBRA density. 
Our upper limit on the magnetic field is in good agreement with  
earlier results of Hillas (\cite{Hill:99}).
Even for $B=0.12$~G intergalactic extinction is needed 
to explain the data above $\sim$10~TeV.
Precision measurements of TeV blazar spectra in the energy region 
from 50~GeV to several hundred GeV, where only negligible intergalactic extinction is expected,
would allow one to select the right value of $(\delta_{\rm j}/B)^{1/2}$ for individual sources.
Such measurements will become feasible with next generation Cherenkov telescope 
systems as HESS and VERITAS and with the satellite instrument GLAST.
Note that it turns out that in our SSC scenarios the energy density of electrons strongly
dominates over the energy density of the magnetic field for all 
admitted values of $B$ and $\delta_{\rm j}$ 
(see also Inoue \& Takahara \cite{Inou:96}).
\subsection{Implications for the intergalactic absorption}
\label{debraresult}
\begin{figure}
\resizebox{\hsize}{!}{\includegraphics{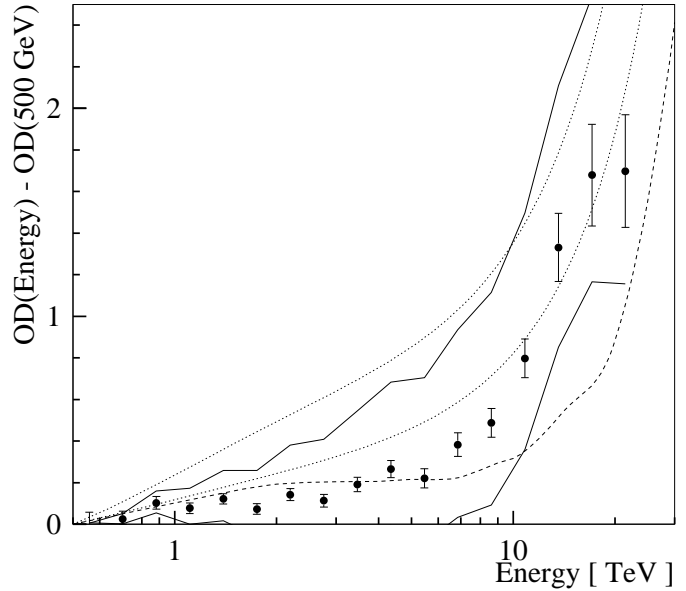}}
\caption
{The increase of the optical depth (OD) for $\gamma$-rays due to 
intergalactic extinction 
as inferred from comparing the observed VHE Mkn 501 spectrum with the 
estimated emitted spectrum (from Fig.~\ref{scen1b}). 
The 1997 Mkn 501 time-averaged spectrum measured by HEGRA has been used and 
the corresponding statistical errors are shown by the vertical error bars. 
Additional uncertainties, arising from the assumed value for the magnetic field $B$ and the
Doppler factor $\delta_{\rm j}$  are shown by the solid lines 
(solid points: $\delta_{\rm j}=25$, $B=0.037$~G,
upper solid line: $\delta_{\rm j}=100$, $B=0.012$~G,
lower solid line: $\delta_{\rm j}=25$, $B=0.12$~G).
The dashed line shows the prediction by Primack et al.\ (\cite{Prim:99})
based on the cosmological model ``LCDM'' with cold dark matter.
The dotted lines show a range of predicted optical depths from 
(Malkan \& Stecker \cite{Malk:98}; Stecker \cite{Stec:99}).
}
\label{od}
\end{figure} 
The increase of the optical depth of the DEBRA above 500~GeV as inferred 
from comparing the model estimate of the emitted VHE 
spectrum with the observed one is shown in Fig.~\ref{od} (solid points). 
The Doppler factor and the magnetic field have been varied as before to illustrate the
effect on the inferred optical depth which amounts here to about 1/2 unit (see 
solid lines). 
As pointed out already, the method primarily gives the dependence of the 
optical depth with energy rather than the absolute amount of absorption.
The result is consistent with a rather energy independent extinction up to energies 
of about 5~TeV. From 5~TeV to 20~TeV the inferred optical depth 
increases by 1 to 2 units.
The dashed line in Fig.~\ref{od} shows the optical depth expected from the 
DEBRA estimates by Primack et al.\ (\cite{Prim:99}) based on galaxy and star 
formation calculations.
The dotted lines in Fig.~\ref{od} show the optical depths for the ``high'' and
``low'' intensity DEBRA models of Malkan \& Stecker (\cite{Malk:98}).
Their results are based on an empirical approach where luminosity dependent 
infrared spectra of galaxies are integrated over their luminosity and 
redshift distributions. 
The figure shows that the increase in optical depth suggested by our 
analysis lies in the range of the theoretical expectations.
The very weak energy dependence of the optical depth up to $\sim$10~TeV
inferred from our analysis agrees well with the predictions of Primack et al..
Our calculations clearly favor the low over the high intensity DEBRA model 
of Malkan \& Stecker. The optical depth predicted by the latter seems to increase
too fast with $\gamma$-ray energy.
\section{Summary}
\label{DISC}
We have analyzed a data sample of simultaneous and nearly simultaneous
X-ray and VHE data. 
The X-ray and VHE observations were performed 
with typical time delays below 6~h.  Since X-ray and the VHE variability is 
found on time scales slower 1/2~day and the spectral indices change typically by 
less than 0.01 units per hour, the measurements are well suited for a meaningful correlation
analysis. For the first time a data base with detailed spectral and temporal information
in the X-ray and VHE bands is available which covers a considerable number
of distinct flares with substantial flux increases and decreases.

The strong variability signatures in both energy bands has been used to 
elucidate several key aspects of the emission activity.
The 3-25~keV spectral index varied on a time scale of weeks from 0.7 to 1.1, 
indicating that the peak of the X-ray SED shifted on this time scale
from the energy region above 25~keV to below several keV and vice versa. 
In Pian et al.\ (\cite{Pian:98b}) a similar shift has been reported from observations
taken in April 1997 and in May 1998 and was attributed to decline of the 
X-ray and VHE $\gamma$-ray activity of the source from 1997 to 1998. The observations presented here
show that the shift actually occured also during the year of increased emission itself.
Furthermore, the time lag between the 3~keV and the 25~keV emission is constrained 
to be smaller than half a day. The data shows an excellent correlation of the 
X-ray fluxes at $\simeq\,25\,$keV with the 2~TeV flux levels.
The data therefore firmly establishes the correlation of the X-ray 
and VHE $\gamma$-ray flux levels of Mkn~501.
A time lag between the 
X-ray and VHE $\gamma$-ray fluxes is constrained to be smaller than one day which is consistent 
with earlier results 
(Catanese et al.\ \cite{Cata:97}; A99a; Aharonian et al.\ \cite{Ahar:99c}).
We find that the spectral variation in X-rays and in the VHE $\gamma$-rays 
could well be of the same magnitude, 
but the accuracy of the VHE measurements does not allow us to draw 
definitive conclusions.

We interpret the data in the framework of SSC models.
Using the constraints on the Doppler factor and the magnetic field
of the emission volume from the variability and correlation properties in the
X-ray and VHE bands we reconstruct the electron spectrum 
from the observed synchrotron radiation and estimate the emitted VHE spectrum.
It turns out that intrinsic source properties, namely a turnover in the electron spectrum,
is expected to cause a turnover of the VHE spectrum just in the TeV energy range. 
This result shows that the VHE measurements give, apart from rather robust 
upper limits on the DEBRA density from general arguments as described 
e.g.\ by Biller et al.\ (\cite{Bill:98}) and A99b, 
reliable information about the intergalactic extinction only
when the emission mechanism is understood in full detail.
Within our model scenario the comparison of observed and predicted VHE spectrum 
gives an optical depth due to intergalactic extinction which is constant
up to energies of about 5~TeV and rises by one to two units from 5~TeV to 20~TeV.
Our future work will focus on a confirmation of the results presented in this paper
by detailed time dependent model studies based on the code by Coppi (\cite{Copp:92}).
\\[3ex]
{\it Acknowledgments}
The analysis of the RXTE data has been possible due to the
High Energy Astrophysics Science Archive Research Center Online Service,
provided by the NASA/Goddard Space Flight Center. We are grateful to E.\ Pian for providing us
with the BeppoSAX data points from April, 1997. H.K.\ and F.A.\ are grateful to 
W.\ Hofmann, J.\ Kirk, and H.\ V\"olk for very fruitful discussions. We thank R.\ Tuffs for
a careful reading of the text. We thank the referee A.M.\ Hillas 
for very useful suggestions.

\end{document}